\documentclass[a4paper,twocolumn]{article}

\usepackage{hyperref}
\usepackage{tikz}
\usetikzlibrary{shapes,arrows,positioning, calc}
\usepackage{graphicx}
\usepackage{authblk}

\usepackage{amsfonts}
\usepackage{amsmath}
\usepackage{mathtools}
\usepackage[space,multidot]{grffile}
\usepackage{subcaption}
\usepackage{booktabs}

\DeclareMathOperator*{\argmax}{arg\,max}

\graphicspath{{./images/pv_jpg2/}}

\newcommand\myscale{0.4}
\newcommand\myhalfscale{0.2}

\author[ ]{Vinicius Pavanelli Vianna}
\author[ ]{Luiz O. Murta Jr.}
\affil[ ]{Department of Physics Applied to Medicine and Biology, Faculty of Philosophy, Science and Languages of Ribeirao Preto, University of Sao Paulo, Ribeirao Preto, SP, Brazil\\}
\affil[ ]{\texttt{•} {\{vnvianna,murta\}@usp.br}}
\title{Analysis of Generalized Entropies in Mutual Information Medical Image Registration}

\begin{document}

\twocolumn[
  \begin{@twocolumnfalse}
    \maketitle
    \begin{abstract}
    Mutual information (MI) is the standard method used in image registration and the most studied one but can diverge and produce wrong results when used in an automated manner. In this study we compared the results of the ITK Mattes MI function, used in 3D Slicer and ITK derived software solutions, and our own MICUDA Shannon and Tsallis MI functions under the translation, rotation and scale transforms in a 3D mathematical space. This comparison allows to understand why registration fails in some circumstances and how to produce a more robust automated algorithm to register medical images. Since our algorithms were designed to use GPU computations we also have a huge gain in speed while improving the quality of registration.\\
    \end{abstract}
  \end{@twocolumnfalse}
]

%\maketitle

%\begin{abstract}
%Mutual information (MI) is the very standard method used in image registration and the most studied one but can diverge and produce wrong results when used in an automated manner. In this study we compared the results of the ITK Mattes MI function, used in 3D Slicer and ITK derived software solutions, and our own MICUDA Shannon and Tsallis MI functions under the translation, rotation and scale transforms in a 3D mathematical space. This comparison allows to understand why the registration fails in some circumstances and how to produce a more robust automated algorithm to register medical images. Since our algorithms were designed to use GPU computations we also have a huge gain in speed while improving the quality of registration.
%\end{abstract}

\section{Introduction}
%In this study we will analyze how generalized entropies changes the aspect of Mutual Information (MI) metrics when used on head MRI with a focus on brain image registration.

Image registration is the alignment of two images in the same geometric space so that structures in images as are overlapped good as possible. Different medical imaging like MRI, fMRI and CT provides different information not only in the coordinate system and resolution but also on identifying different kinds of morphological and/or functional structures. To better use those different information one need to register images so that they overlap and align making possible to use the best information from each kind of imaging technique for a better clinical diagnosis. Other medical usage of registration is to compare multiple different patients. This multiple patients comparison generate medical atlas like the MNI-ICBM\cite{McKinstry2010,Collins} providing a common ground to morphological structures.

The standard technique in medical image registration is the Mutual Information (MI) introduced by Shannon \cite{shannon1963mathematical} and used first by Viola and Collignon \cite{Viola1997,Collignon97}, shown here in the Shannon entropy equations (\ref{eqn:eqshannon}) and the MI equation (\ref{eqn:eqmi}).

As MI only provides a measurement of how good is the alignment between two images we need to use an optimizer that will change the geometric transformation parameters trying to maximize (or minimize) the metric function to achieve the goal of image registration.

\begin{align}
\label{eqn:eqshannon}
\begin{split}
H(x) &\equiv -\sum p(x) \; \log \, p(x)
\\
H(x,y) &\equiv -\sum p(x,y) \; \log \, p(x,y)
\end{split}
\end{align}

\begin{equation}
\label{eqn:eqmi}
I(A;B) = H(A) + H(B) - H(A,B)
\end{equation}

\begin{figure}
   \caption{Registration Block Diagram}
   \label{diagrama_bloco}
\resizebox{0.5\textwidth}{!}{   
\tikzstyle{block} = [draw, fill=blue!25, rectangle, 
    minimum height=3em,  font=\footnotesize]
%\tikzstyle{sum} = [draw, fill=blue!20, circle, node distance=1cm]
\tikzstyle{input} = [coordinate]
\tikzstyle{output} = [coordinate]
%\tikzstyle{pinstyle} = [pin edge={to-,thin,black}]

%\begin{figure}
%   \caption{\label{diagrama_bloco}Block Diagram}
%\begin{tikzpicture}[auto, node distance=1cm,>=latex']
\begin{tikzpicture}[
auto, 
ultra thick, 
>=latex',
%transform canvas={scale=0.8}
]
\node [block, minimum height=2cm] (metrica) {Metric};
\node [coordinate, above=5mm of metrica.west] (metrica_fixa) {};
\node [coordinate, below=5mm of metrica.west] (metrica_movel) {};
% \node [input, left=of metrica.135] (input_fixa) {Fixa};
\node [block, right=of metrica] (otimizador) {Optimizer};
\node [output, right=2cm of otimizador] (saida) {};
%\node [block, left=2cm of metrica_movel] (interpolador) {Interpolator};
\node [block, left=2cm of metrica_movel
, text width=6em, align=center
] (tg) {Transform};

\node [input, left=2cm of tg] (input_movel) {};
%\node [input, left=of metrica_fixa] (aux_fixa) {};
\node [input] (input_fixa) at (metrica_fixa -| input_movel) {Fixed};
\node [coordinate] (aux_fixa) at (input_fixa -| tg.west) {};
\node [coordinate] (aux_fixa_interpolador) at (input_fixa -| tg.east) {};
\node [coordinate] (aux_saida) at ( $ (otimizador)!0.5!(saida) $ ) {};
\node [coordinate, below=of tg] (feed_tg) {};
\node [coordinate] (feed_saida) at (feed_tg -| aux_saida) {};
%\node at (input_fixa) [above = 1mm of input_fixa] {Fixa};
\draw[-] (input_fixa) -- node {Fixed} (aux_fixa);
\draw[-] (aux_fixa) -- (aux_fixa_interpolador);
\draw[->] (aux_fixa_interpolador) -- node {$u(x)$} (metrica_fixa);

%\draw[->] (tg) -- (interpolador);
\draw[->] (tg) -- node {$v(T(x))$} (metrica_movel);
\draw[->] (input_movel) -- node {Moving} (tg);
\draw[->] (metrica) -- node {$F_m(\cdot)$}(otimizador);
\draw[->] (otimizador) -- node {$\hat{T}$} (saida);
\draw[-] (aux_saida) -- (feed_saida);
\draw[-] (feed_saida) -- node {$T$} (feed_tg);
\draw[->] (feed_tg) -- (tg.south);
\end{tikzpicture}
%\end{figure}
}
\end{figure}
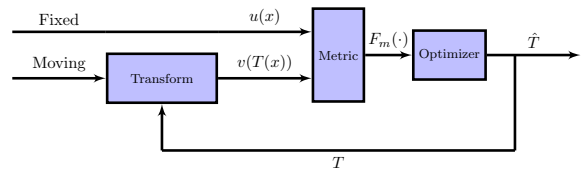

Figure \ref{diagrama_bloco} shows the basic block diagram of a image registration algorithm, the fixed and moving ($u(x)$ and $v(x)$) images are the inputs on the left with the moving image passing by a \texttt{Transform} block that receive parameters ($T$) from the \texttt{Optimizer} block, then the fixed and the transformed moving (now $v(T(x))$) image goes to the \texttt{Metric} function ($F_m(\cdot)$) that measures how good is the current transform in solving the registration. The \texttt{Optimizer} receives this measure information and updates the parameters sent to the \texttt{Transform} block until it reaches a value that shows a match on the registration ($\hat{T}$). This value can be the maximum or minimum of the \texttt{Metric} depending on the type of the function $F_m(\cdot)$ used. Mathematically this algorithm can be represented by the equation \ref{eq_argmax}.

\begin{equation}
\label{eq_argmax}
			 \hat{T} = \argmax_{T} F_{m}(u(x), v(T(x))  
\end{equation}

A very robust and still used optimizer is the gradient descent (created in 1847 by Cauchy \cite{Cauchy1847}) that evolved to the stochastic gradient descent method \cite{Kiefer1952a}. These methods use the gradient of the metric function to make steps in guessing of the best transformation parameters so each consecutive guess is better than the previous one. This guessing goes on until the global maximum is achieved and we find our solution or the optimizer locks in a local maximum that is not our solution and can't be further improved.

To make the registration process more robust we can use an optimizer that can deal with those various local maxima such as genetic algorithms or even exhaustive search over all the parameters, or we can use a metric that only have one maximum point. Our quest is to search for such metric function.

We believe that when using an information technique we need to use all information available and with recent computational improvements we can do that without the need to wait several hours for a single measurement. In this way our study goes beyond the simple change to a generalized entropy to measure how the information quantity provided to the MI metric affect it image registration capability.

We also provided an in-depth study of the generalized entropies, such as Tsallis entropy, when used for image registration in the geometric transforms of translation, rotation and scale. Those transforms forms the base of the affine family along the skew transform and cover mostly all simple image registration. When there is need for a more complex registration one can use the B-Spline transform.

\section{Methodology}
We used brain images kindly provided by the Human Connectome Project\cite{Glasser2013,VanEssen2011} in their HCP Young Adult study. The full image protocols can be found at their website\footnote{\url{http://protocols.humanconnectome.org/HCP/3T/imaging-protocols.html}}. In this study we only used the 3T pre-processed structural images.

Our main comparison was with the ITK Mattes Mutual Information technique\cite{Mattes2001a} provided by the ITK software\cite{Yoo2002} and used in the BRAINSFit\cite{brainsfit} module of the 3DSlicer software\footnote{\url{http://www.slicer.org/}}.

To allow visualization and comparison of the metric we generated 3D images by varying the parameters of a single family of transform (i.e. when we see the translation 3D image in the X axis we can see how the metric changes when the image was translated along the X axis). This is equivalent of an exhaustive search over the space, so what we did was to map with the MI function all the space in a finite range and build a 3D image with this map.

This separation of geometric transform in translation, rotation and scale was needed since a normal image registration can involve 12 or even more degrees of freedom (or dimensions) and visualization of multidimensional data like those can be very difficult. Separating the transforms we could generate simple 3D result images that can be easily analyzed.

Another benefit of separating is that we can process the parameters space in a finer way. To use 51 individuals positions of translations in 3 dimensions (i.e. using a translation from $-25mm$ to $25mm$ in every direction) can result in $51^3 = 132651$ measurements. Using 12 dimensions and only 3 individual parameter (i.e. a minus parameter, a plus parameter and zero) in each dimension or direction we have $3^{12} = 531441$ or $4\times$ the 3 dimension approach and pretty much no idea of how each dimension affect the output of the metric! This will become more clear in the results section.

Making all those measurements can be hard and time consuming. To improve the performance we migrate our initial code to a GPU based one using NVIDIA CUDA toolkit\footnote{\url{https://developer.nvidia.com/cuda-zone}}. This allowed a huge gain in performance with a non expensive investment in hardware making our solution more viable to further research and clinical use. All the data here presented were made using a NVIDIA GTX 1060 and used less than 2GB of GPU memory allowing it to run on less powerful GPU hardware if needed.

\section{Results}

% This should be methodology??

As said before our results are the mathematical image of the MI function of two images. We made a 3D cube with each point representing the result of the MI function analyzed with a transformation related to that point coordinates. So if we check a point in the cube at position $(10,10,150)$ that values represents the MI of two images with the moving image transformed with the $(10,10,150)$ parameters. If we take the transformation to be a translation that point would be the MI result of two images with the moving image translated by $10mm$ at axis X, $10mm$ at axis Y and $150mm$ at axis Z.

In this way we can understand what happens inside the registration algorithm. The gradient descent will see those MI results and try to register the image with the minimum value it can reach. The gold standard in our case is the central point $(0,0,0)$ in that the moving image would have not be transformed in any way since initially our images was the same or already registered.

Another point to remember is that different entropies will give different results, so some methods will give the center point as a local maximum as others will give as a local minimum. This is specially true in the Tsallis entropy when we go from $q<1.0$ to $q>1.0$ and will be noted on the following figures as some have the center in blue and some in brown reflecting this change.

\subsection{Translation}

\begin{figure*}
\centering
\subcaptionbox{}{\includegraphics[scale=\myscale]{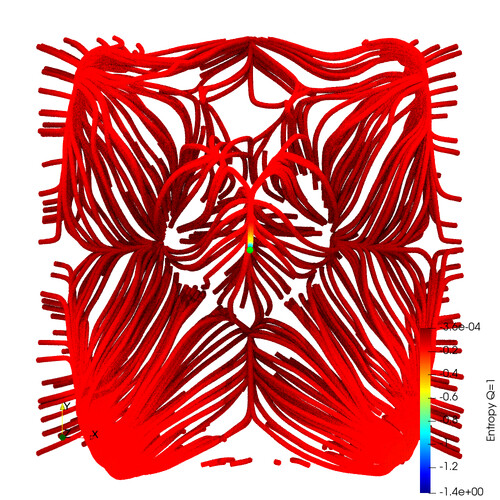}}%
\hfill
\subcaptionbox{}{\includegraphics[scale=\myscale]{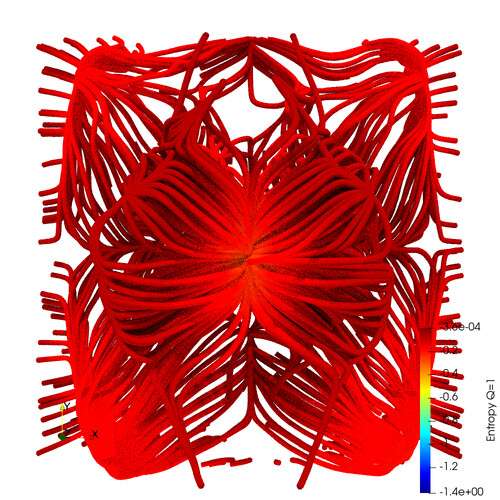}}%
\hfill
\subcaptionbox{}{\includegraphics[scale=\myscale]{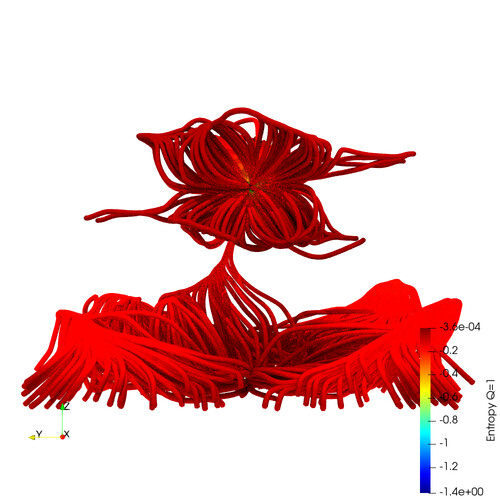}}%
\hfill
\subcaptionbox{}{\includegraphics[scale=\myscale]{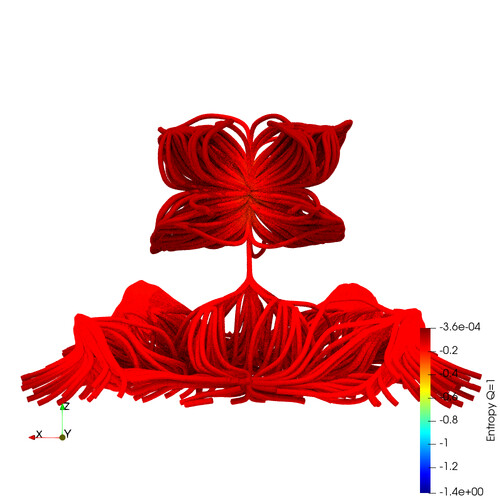}}%
\caption{\label{path_mattes}Paths of the gradient divergent algorithm along the MI of the same image with only translation transform and the ITK Mattes algorithm: (a) the paths emerging from the $z=-150$ plane on the back viewing from the $z$ axis (b) same as before with the paths along the center added (c) same as before from the $x$ axis (d) same as before from the $y$ axis}
\end{figure*}

We start by analyzing the paths the gradient descent algorithm will take when registering two images using different MI functions. Figure \ref{path_mattes} shows those paths when we use the ITK Mattes function with the same image on both fixed and moving. The first image (a) is the paths emerging from the lower plane ($z=-150mm$), it can be seen five different regions with a central region converging to the central point (our gold standard and correct point) and four regions on the corners converging to other points not related to our goal. The second image (b) we added the central block of paths, all connected to our central point and from where the ITK Mattes can correctly register the images. 

As its difficult to see a 3D function we added two other views of the (b) images, as the view from the $X$ axis in (c) and from the $Y$ axis in (d). From those images it is very clear that beside that central region in (a) all other regions will not connect to the central point as there is no path between then opposed to the central region where there is always a path between it and the $(0,0,0)$ point as can now be seen in (c) and (d). The $(0,0,0)$ or central point is the very center of the upper structure visualized in (c) and (d) and added in (b).

\begin{figure*}
\centering
\subcaptionbox{}{\includegraphics[scale=\myscale]{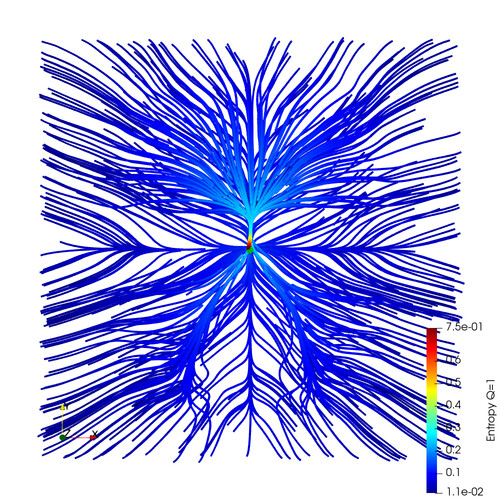}}%
\hfill
\subcaptionbox{}{\includegraphics[scale=\myscale]{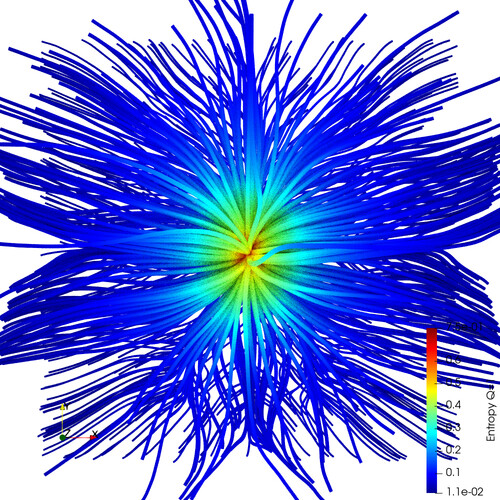}}%
\hfill
\subcaptionbox{}{\includegraphics[scale=\myscale]{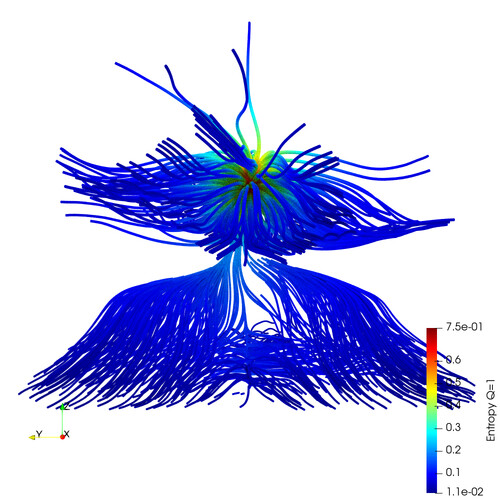}}%
\hfill
\subcaptionbox{}{\includegraphics[scale=\myscale]{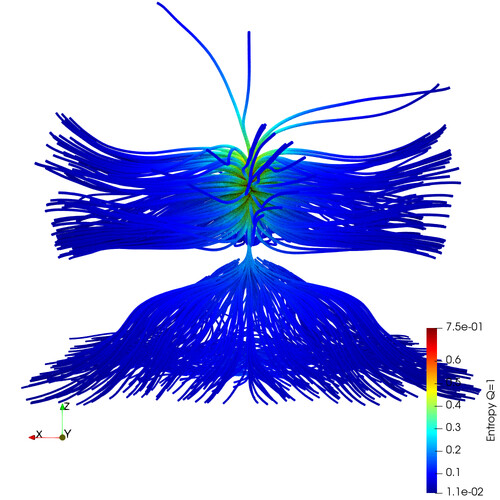}}%
\caption{\label{path_shannon}Paths of the gradient divergent algorithm along the MI of the same image with only translation transform and the MICUDA algorithm with Shannon entropy: (a) the paths emerging from the $z=-150$ plane on the back viewing from the $z$ axis (b) same as before with the paths along the center added (c) same as before from the $x$ axis (d) same as before from the $y$ axis}
\end{figure*}

The paths of the gradient descent algorithm changes dramatically when we use the MICUDA with the Shannon entropy in Figure \ref{path_shannon}. Now we can see in the first image (a) almost all regions converging to the central point, there is a small region that will not converge on the horizontal extremes that will converge to a local minimum under the central point. On the second image (b) the centrals paths added are much bigger than with ITK Mattes and now reach all the cube boundary. On the third image (c) we can see better that region of (a) that will not converge, showed on the bottom center, a smaller region than the one in Figure \ref{path_mattes}. On the fourth image (d) we can check that on the upper part of (a) now shown better here we have full convergence to the central point.

It should be noted that the central paths in Figures \ref{path_mattes} and \ref{path_shannon} don't show all the paths related to the central point but only those with bigger gradients, so even when the central point don't have visible paths to all regions on the images there can be paths to that regions. But regions with paths shown can't have other paths that diverge from those. So when we see a path going to a local minimum (not the center point) on those images they are really there guiding our Gradient Descent algorithm the wrong way.

\begin{figure*}
\centering
\subcaptionbox{}{\includegraphics[scale=\myscale]{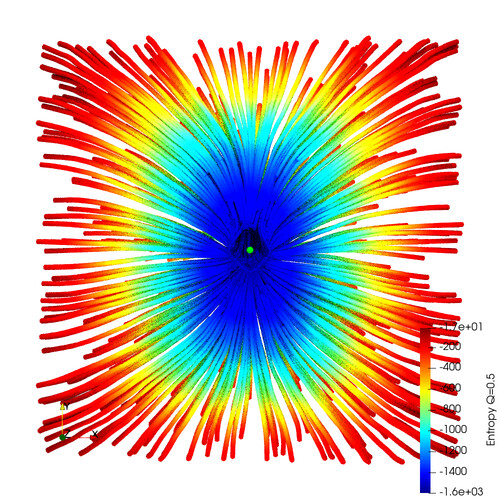}}%
\hfill
\subcaptionbox{}{\includegraphics[scale=\myscale]{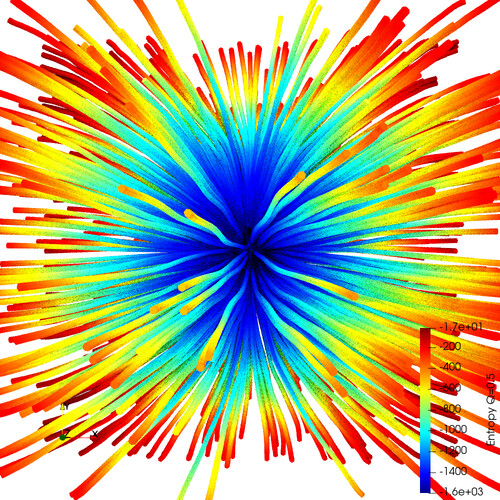}}%
\hfill
\subcaptionbox{}{\includegraphics[scale=\myscale]{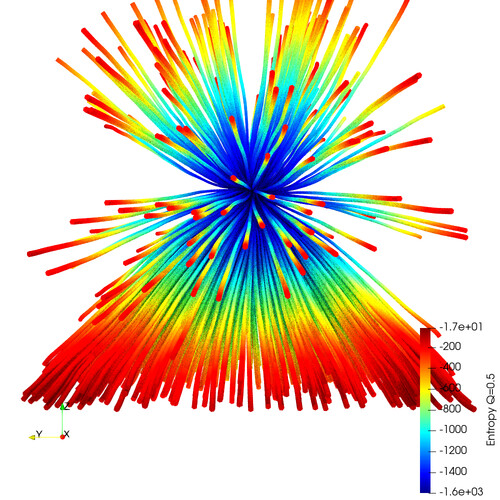}}%
\hfill
\subcaptionbox{}{\includegraphics[scale=\myscale]{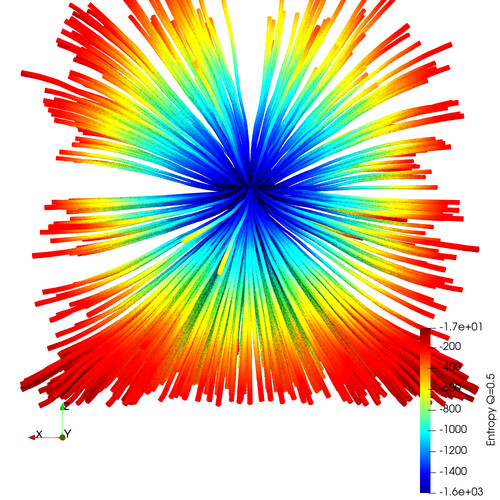}}%
\caption{\label{path_q_05}Paths of the gradient divergent algorithm along the MI of the same image with only translation transform and the MICUDA algorithm with Tsallis entropy and $q=0.5$: (a) the paths emerging from the $z=-150$ plane on the back viewing from the $z$ axis (b) same as before with the paths along the center added (c) same as before from the $x$ axis (d) same as before from the $y$ axis}
\end{figure*} 

\begin{figure*}
\centering
\subcaptionbox{}{\includegraphics[scale=\myscale]{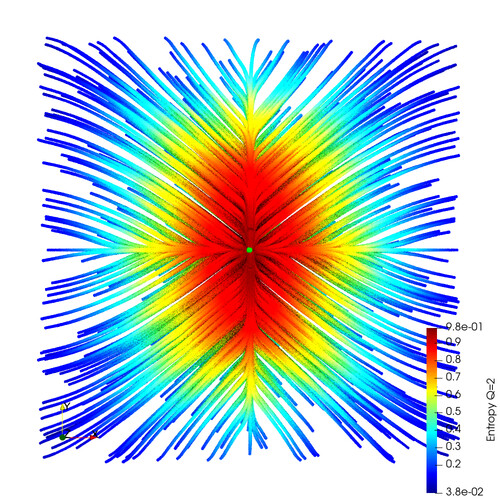}}%
\hfill
\subcaptionbox{}{\includegraphics[scale=\myscale]{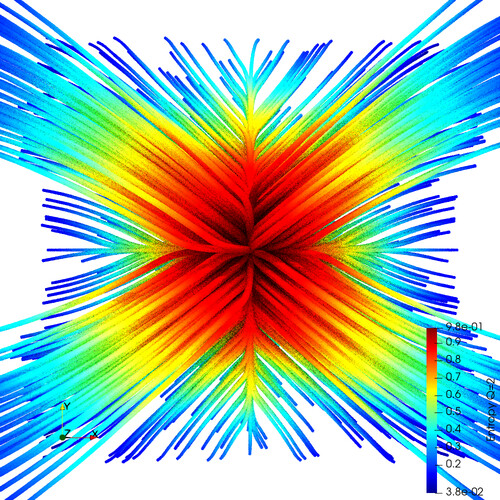}}%
\hfill
\subcaptionbox{}{\includegraphics[scale=\myscale]{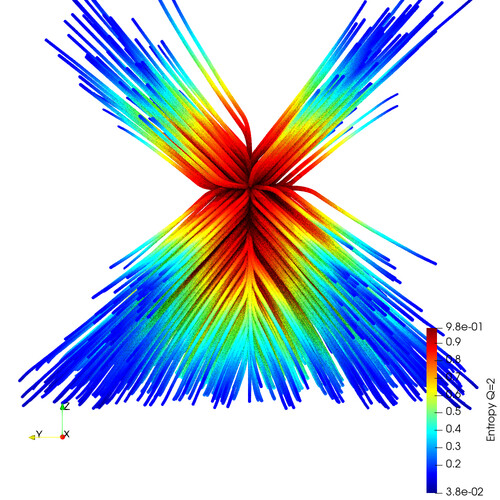}}%
\hfill
\subcaptionbox{}{\includegraphics[scale=\myscale]{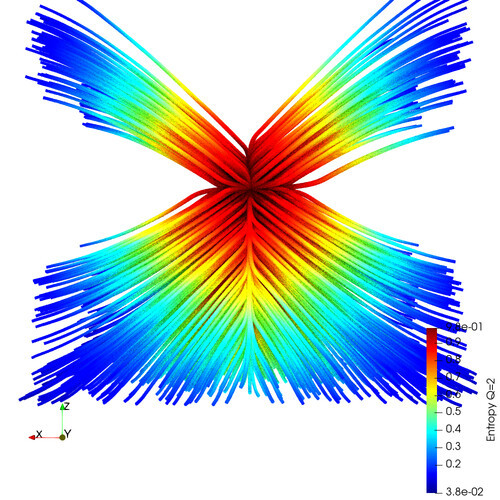}}%

\caption{\label{path_q_20}Paths of the gradient divergent algorithm along the MI of the same image with only translation transform and the MICUDA algorithm with Tsallis entropy and $q=2.0$: (a) the paths emerging from the $z=-150$ plane on the back viewing from the $z$ axis (b) same as before with the paths along the center added (c) same as before from the $x$ axis (d) same as before from the $y$ axis}
\end{figure*}

Next we look into the effects of Tsallis entropy on the MI function. Starting with Figure \ref{path_q_05} that have a $q$ value of $0.5$ we can see on the first image (a) a very nice convergence almost to the central point since the paths stop very near it forming some kind of closed loop (that will be discussed later). On the other images we can see that the central paths take almost all regions of the cube and the lower $z=-150mm$ plane paths don't form the ``bell'' shaped of former figures but now have a more direct way to the central point. That direct way is better since our gradient descent won't have to make corrections on it's course to the central point with each intermediate point having a gradient pointing almost equal to the former, in a quasi-direct line.

If we push the Tsallis $q$ value to $2.0$ we have the Figure \ref{path_q_20} showing in the first image (a) a full convergence to the central point. In the later images (b-d) we can see the same direct paths to the central point as the former Tsallis figure but now the central paths is not distributed as before but having much bigger gradients to the corners.

\begin{figure*}
\centering
\subcaptionbox{}{\includegraphics[scale=\myscale]{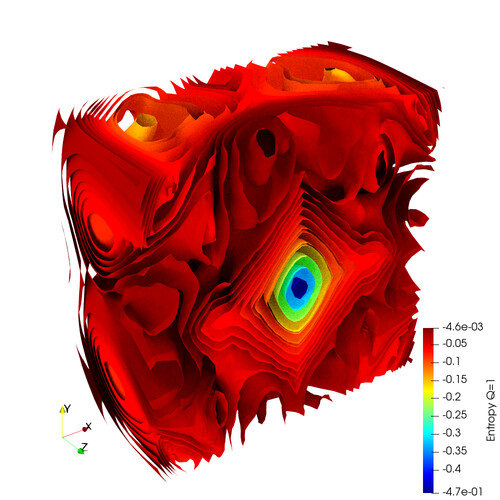}}%
\hfill
\subcaptionbox{}{\includegraphics[scale=\myscale]{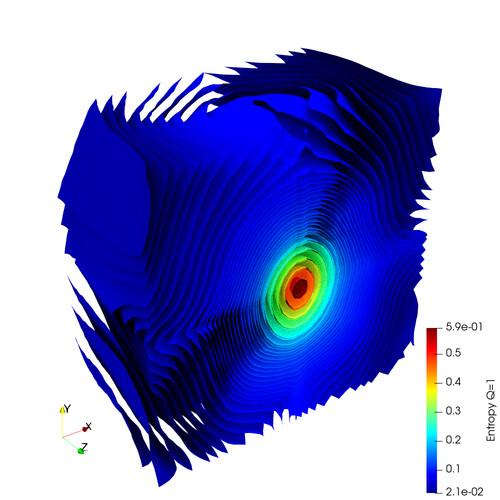}}%
\hfill
\subcaptionbox{}{\includegraphics[scale=\myscale]{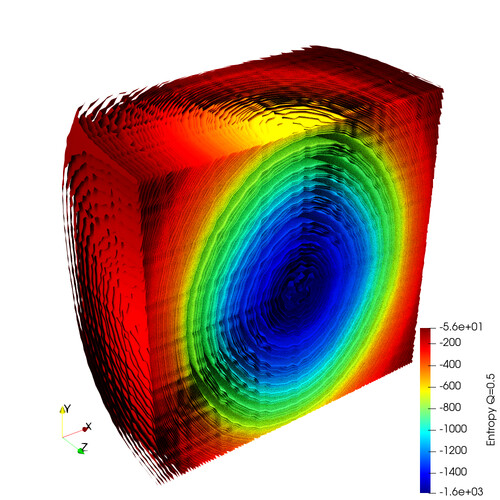}}%
\hfill
\subcaptionbox{}{\includegraphics[scale=\myscale]{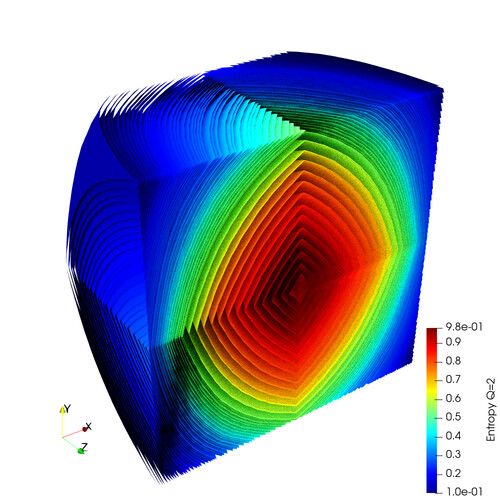}}%
\caption{\label{contour}Isosurfaces from the MI of the same image with only translation transform with multiple methods: (a) ITK Mattes (b) MICUDA Shannon (c) MICUDA Tsallis $q=0.5$ (d) MICUDA Tsallis $q=2.0$}
\end{figure*}

Now that we know the paths the gradient descent we can begin to analyze the contour isosurfaces of the MI function keeping in mind that the gradient paths are always perpendicular (or normal) to those surfaces since they flow along the gradient field of the function and isosurfaces are surfaces were the field have the same value. In Figure \ref{contour} the first image (a) shows the ITK Mattes MI function. It's very clear the big mess on the periphery that will guide our gradient descent algorithm the wrong way. Even paths that can flow to the center point will be disturbed with detours caused by those mess. In more technical terms we say that the function have multiple local minima (or maxima). To the gradient descent algorithm those local minima seems to be the correct solution since the algorithm can't see all the mathematical space of the MI function as we are seeing and can't know that we have a center point that is much lower than those minima. This is the fundamental problem with the ITK Mattes registration in translation transforms.

On the other images (Figure \ref{contour} b-d) we can see that our algorithm provides a much cleaner mathematical space for registration. The second image (b) shows the MICUDA with Shannon entropy were we can see a much nicer local minima at the center point, the only problem here is the surfaces not providing a direct path to the center point so we will have some curves but as seen before we can register almost all the space. The third image (c) shows the Tsallis entropy with $q=0.5$ were we have very nice spherical surfaces but they are not smooth as we want. Some irregularities emerge from the solution and those explain the paths in Figure \ref{path_q_05} (a) not converging to the central point but to a space near it. If we pay attention to the very center of the surfaces we can see they start to look more irregular towards the center, stopping the paths to flow to it. On the fourth image (d) we see very smooth surfaces, not very spherical since they have some corners but the smooth provides a direct path to the center point, explaining the difference between Figures \ref{path_q_05} and \ref{path_q_20}. Now the paths can flow to the center point and we have a solution to our problem, at least in the translation transformation.

\subsection{Rotation}

\begin{figure*}
\centering
\subcaptionbox{}{\includegraphics[scale=\myscale]{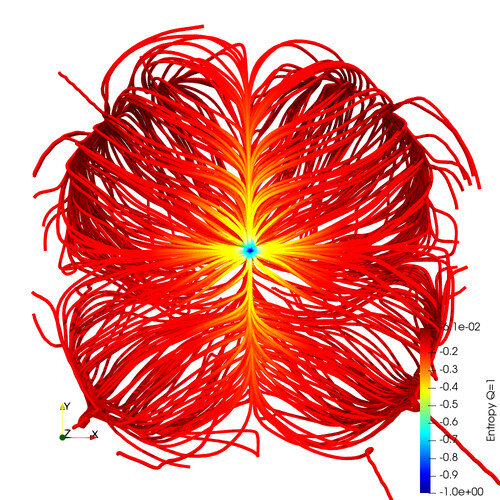}}%
\hfill
\subcaptionbox{}{\includegraphics[scale=\myscale]{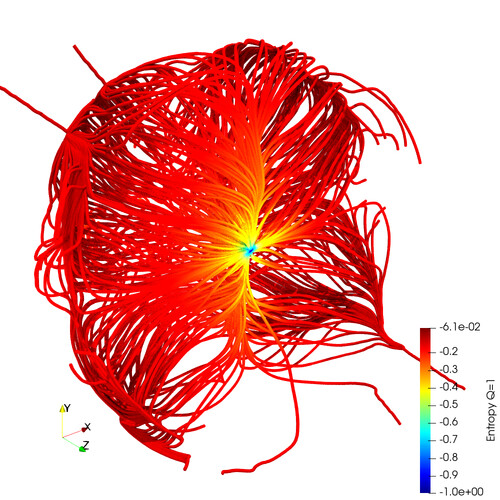}}%
\hfill
\subcaptionbox{}{\includegraphics[scale=\myscale]{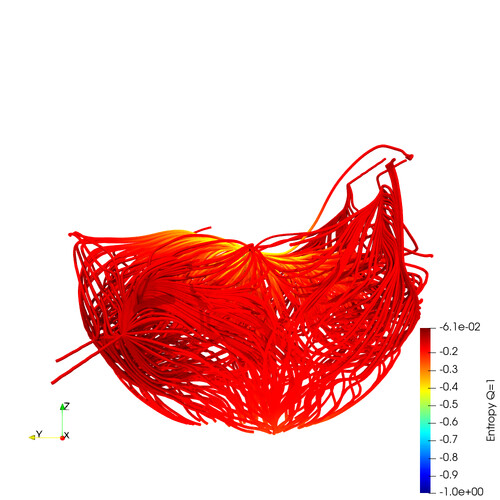}}%
\hfill
\subcaptionbox{}{\includegraphics[scale=\myscale]{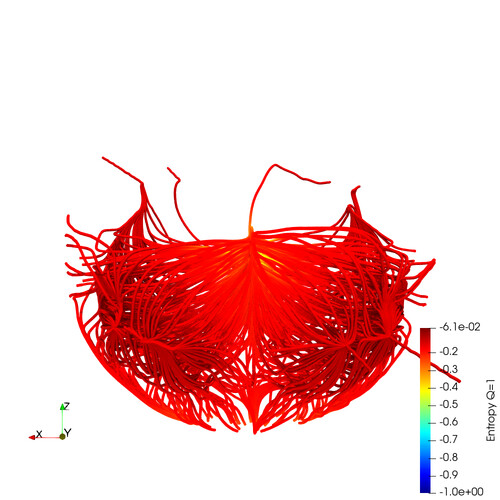}}%
\caption{\label{path_rotation_mattes}Paths of the gradient divergent algorithm along the MI of the same image with only rotation transform and the ITK Mattes algorithm: (a) the paths emerging from the $z<0$ hemisphere on the back viewing from the $z$ axis (b) same as before from different angle (c) same as before from the $x$ axis (d) same as before from the $y$ axis}
\end{figure*}

The rotation transform shows to be more difficult to solve than translation as we can see in the Figure \ref{path_rotation_mattes}. Since all points outside the sphere with $radius = 1$ are mapped to the sphere surface we only show this sphere volume here. It can be seen on the first image (a) that most paths converge to the center point but we have a region that converge to other local maxima on the sphere periphery. This regions is shown on the middle horizontal line of the first image (a), the left center of second image (b) and bottom of the two last images (c and d). On this colormap the gradient flows from the brown to yellow to blue. Noting those color flows we can see two connected graphs with only one converging to the central point. Those two graphs are more clear in the last two images, since the central point connected is on top and the divergent one is on bottom.

\begin{figure*}
\centering
\subcaptionbox{}{\includegraphics[scale=\myscale]{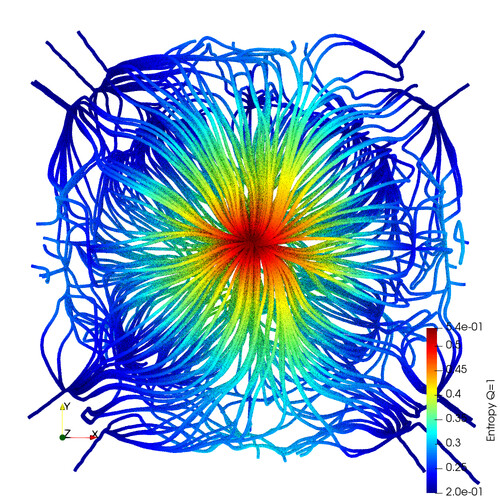}}%
\hfill
\subcaptionbox{}{\includegraphics[scale=\myscale]{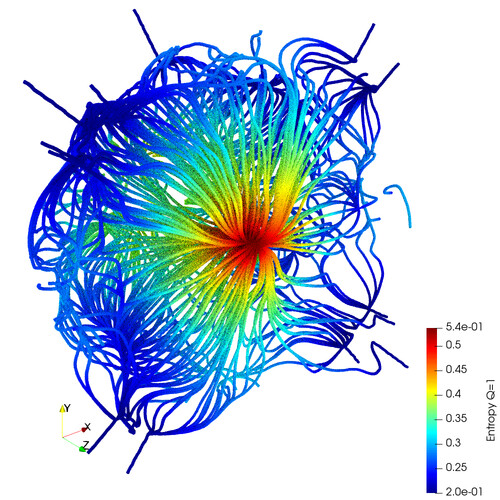}}%
\hfill
\subcaptionbox{}{\includegraphics[scale=\myscale]{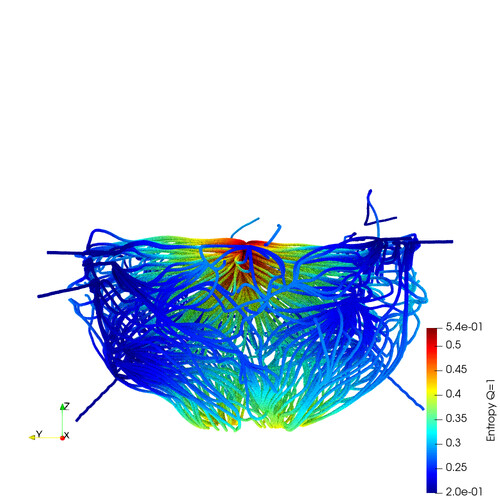}}%
\hfill
\subcaptionbox{}{\includegraphics[scale=\myscale]{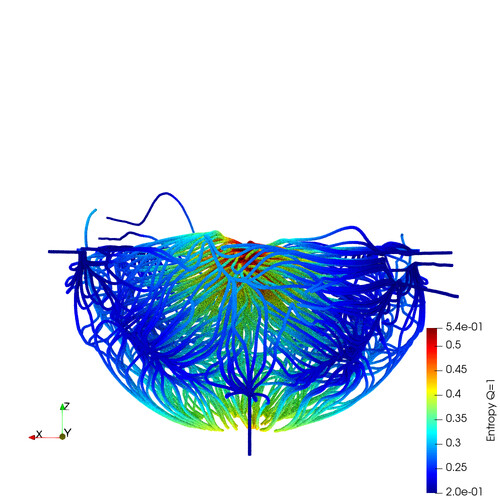}}%
\caption{\label{path_rotation_shannon}Paths of the gradient divergent algorithm along the MI of the same image with only rotation transform and the MICUDA Shannon ($q=1.0$) algorithm: (a) the paths emerging from the $z<0$ hemisphere on the back viewing from the $z$ axis (b) same as before from different angle (c) same as before from the $x$ axis (d) same as before from the $y$ axis}
\end{figure*}

The change from ITK Mattes to MICUDA Shannon (Figure \ref{path_rotation_shannon}) don't show much improvement of the registration capability. The gradient is a little different as can be seen on the first image (a) of both methods. In Mattes the lines converge to central paths along the horizontal or vertical axis ($X$ and $Y$), and in the Shannon method the lines take their own path in a more direct way not related to the axis lines.

\begin{figure*}
\centering
\subcaptionbox{}{\includegraphics[scale=\myscale]{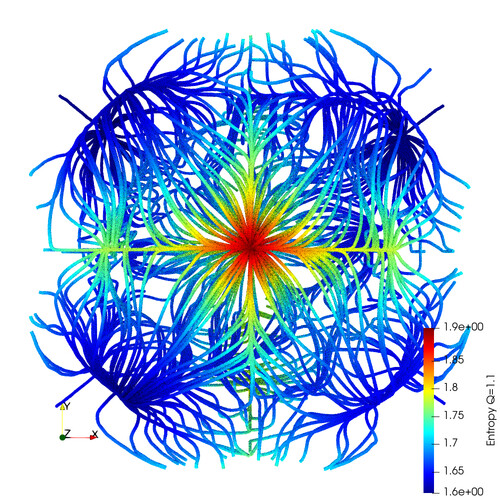}}%
\hfill
\subcaptionbox{}{\includegraphics[scale=\myscale]{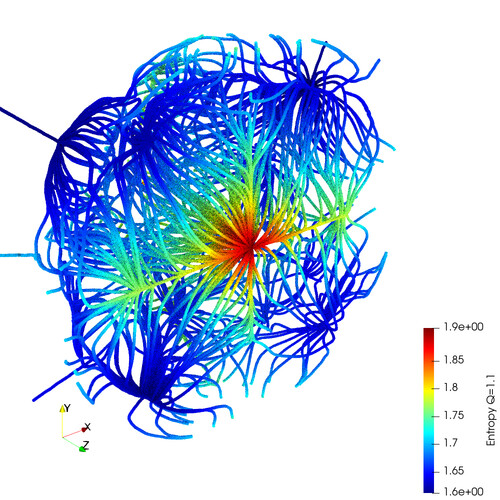}}%
\hfill
\subcaptionbox{}{\includegraphics[scale=\myscale]{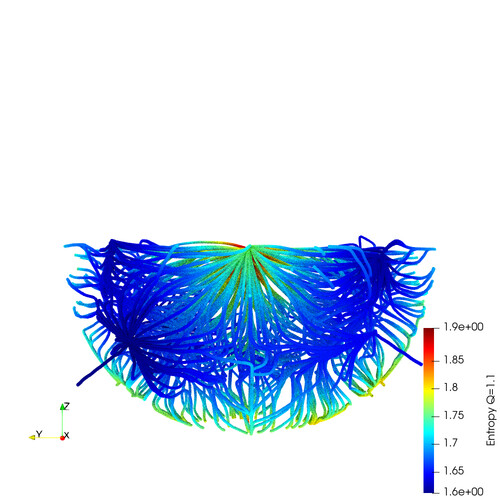}}%
\hfill
\subcaptionbox{}{\includegraphics[scale=\myscale]{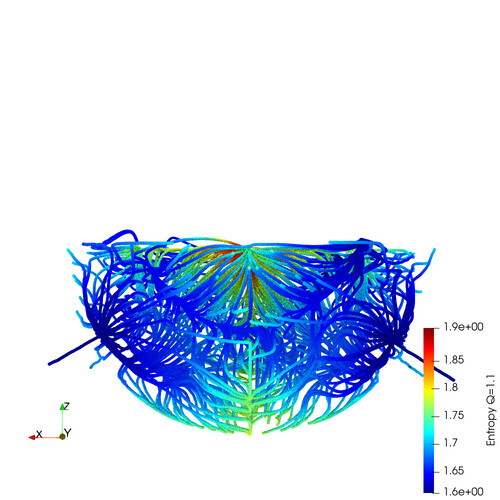}}%
\caption{\label{path_rotation_tsallis_11}Paths of the gradient divergent algorithm along the MI of the same image with only rotation transform and the MICUDA Tsallis ($q=1.1$) algorithm: (a) the paths emerging from the $z<0$ hemisphere on the back viewing from the $z$ axis (b) same as before from different angle (c) same as before from the $x$ axis (d) same as before from the $y$ axis}
\end{figure*}

\begin{figure*}
\centering
\subcaptionbox{}{\includegraphics[scale=\myscale]{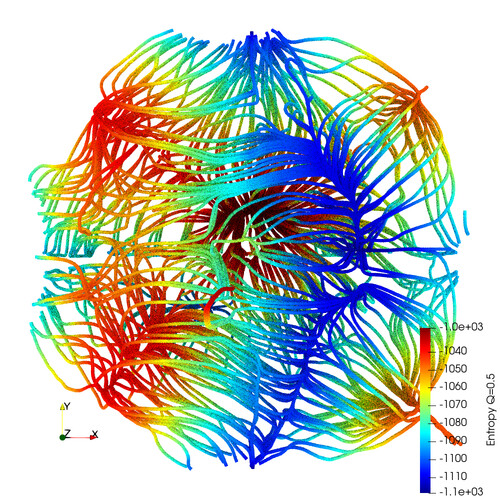}}%
\hfill
\subcaptionbox{}{\includegraphics[scale=\myscale]{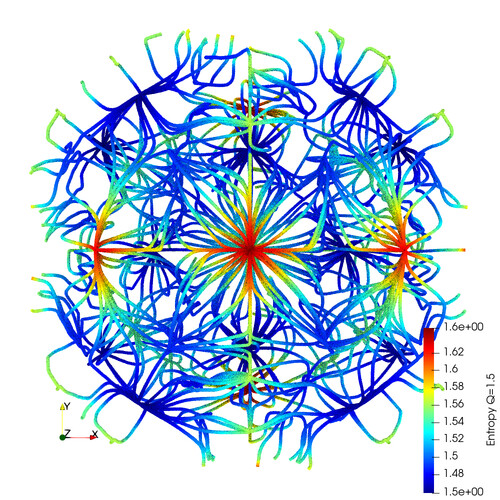}}%
\hfill
\subcaptionbox{}{\includegraphics[scale=\myscale]{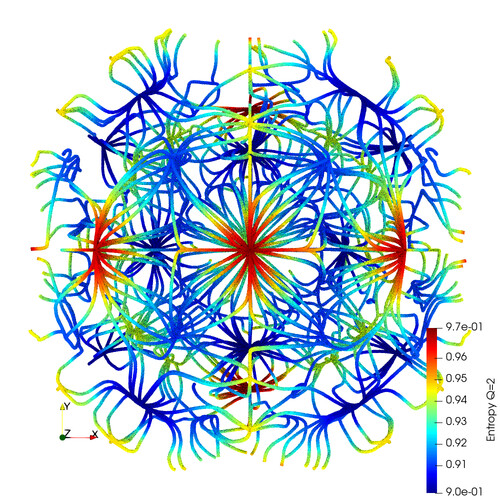}}%
\caption{\label{path_rotation_tsallis_broken}Paths of the gradient divergent algorithm along the MI of the same image with only rotation transform and the MICUDA Tsallis algorithm: (a) $q=0.5$ (b) $q=1.5$ (c) $q=2.0$}
\end{figure*}

The best results using Tsallis came from $q=1.1$ (Figure \ref{path_rotation_tsallis_11}), it's very similar to the Shannon field but we have a little more capture range in some parts of the space. Values above ($q>1.1$) have some local maxima on the horizontal axis that will lower the capture range and values below Shannon ($q<1.0$) have a local maximum displaced from the center point and some other local maxima depending on the values (Figure \ref{path_rotation_tsallis_broken}).

\begin{figure*}
\centering
\subcaptionbox{}{\includegraphics[scale=\myscale]{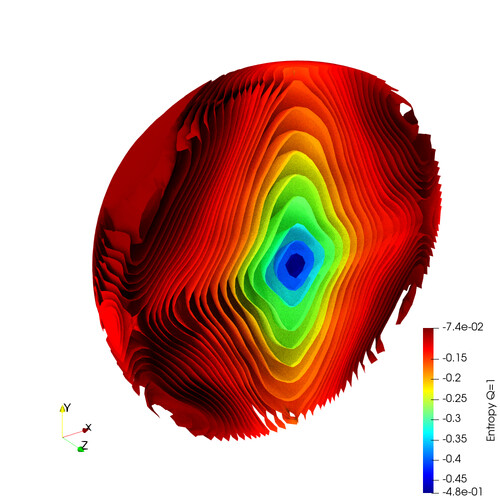}}%
\hfill
\subcaptionbox{}{\includegraphics[scale=\myscale]{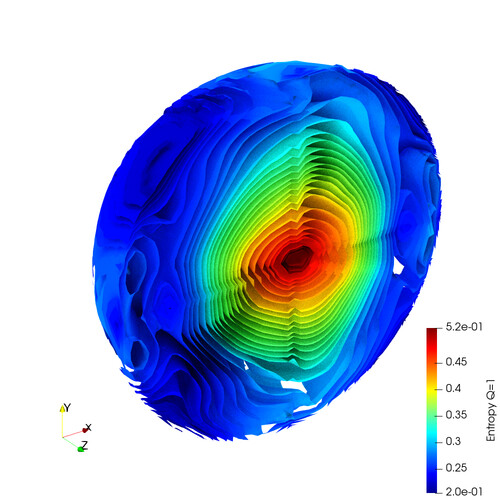}}%
\hfill
\subcaptionbox{}{\includegraphics[scale=\myscale]{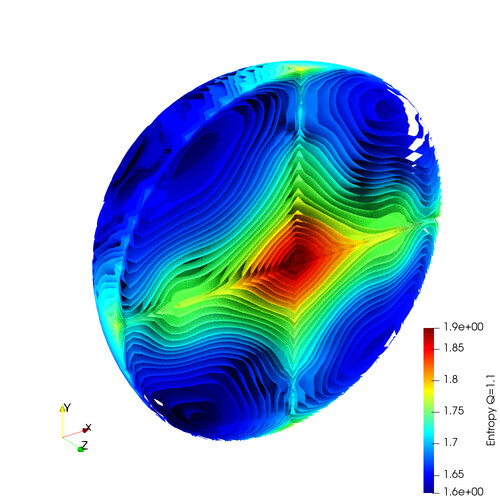}}%
\hfill
\subcaptionbox{}{\includegraphics[scale=\myscale]{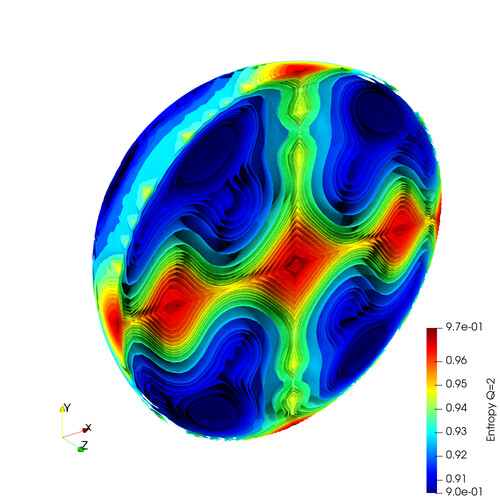}}%
\hfill
\subcaptionbox{}{\includegraphics[scale=\myscale]{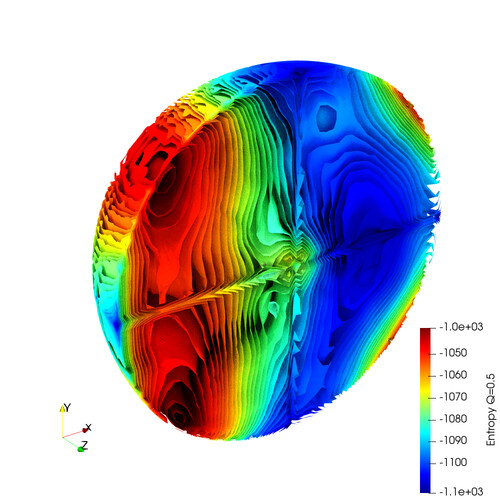}}%
\caption{\label{contour_rotation}Isosurfaces from the MI of the same image with only rotation transform with multiple methods: (a) ITK Mattes (b) MICUDA Shannon (c) MICUDA Tsallis $q=1.1$ (d) MICUDA Tsallis $q=2.0$ (e) MICUDA Tsallis $q=0.5$}
\end{figure*}

The contour isosurfaces of the rotation transform are shown in Figure \ref{contour_rotation}. In the first image we have the ITK Mattes function with an almost full range of capture with the exception of the diagonal corners of the figure, where we have some small local maxima. Almost all paths flow to the center point as we want. In the second image we have the MICUDA Shannon function very similar to the ITK Mattes, a nice range but now the local maxima are in the extreme horizontal and vertical of the image where we can see some yellow surfaces appearing. In the third image we have the MICUDA Tsallis function with $q=1.1$. Now we have a very definite and strong gradient field from the diagonal corners to the center point but we also have the periphery of the sphere not connecting to the center point. In other words the periphery of the sphere diverges and won't register in this situation. The fourth image have the MICUDA Tsallis with $q=2.0$ and the interesting feature is two strong local maxima showing in the horizontal center line and some small ones on the vertical center line. Those local maxima start to emerge from $q>1.1$ and will break our registration by forcing the image to register with a rotation angle. The fifth image is the MICUDA Tsallis with $q=0.5$ and shows a very disturbing gradient field to our registration process with the local maxima on the left diagonal corners and no local maxima on the center point. In this $q=0.5$ even perfect registered images will be rotated to a wrong angle by this function so it's completely useless for rotation transform. This problem appears on all $q<1.0$ researched by us on rotation.

% TODO
% should discuss this bigger gradients effect on the registration!!!

\subsection{Scale}

\begin{figure*}
\centering
\subcaptionbox{}{\includegraphics[scale=\myscale]{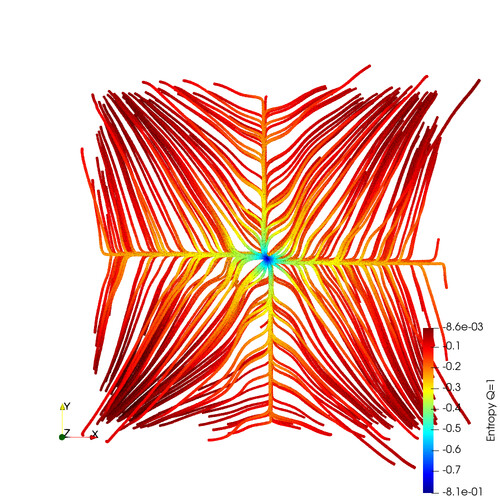}}%
\hfill
\subcaptionbox{}{\includegraphics[scale=\myscale]{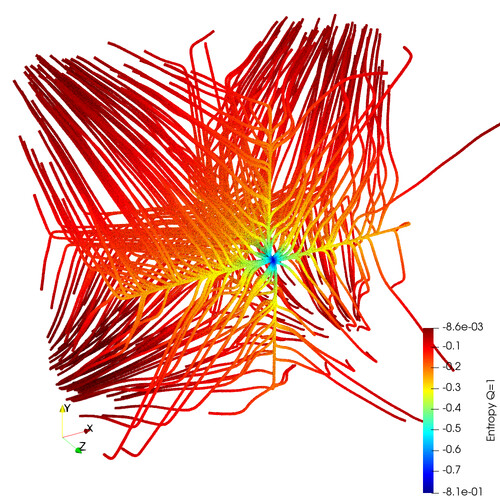}}%
\hfill
\subcaptionbox{}{\includegraphics[scale=\myscale]{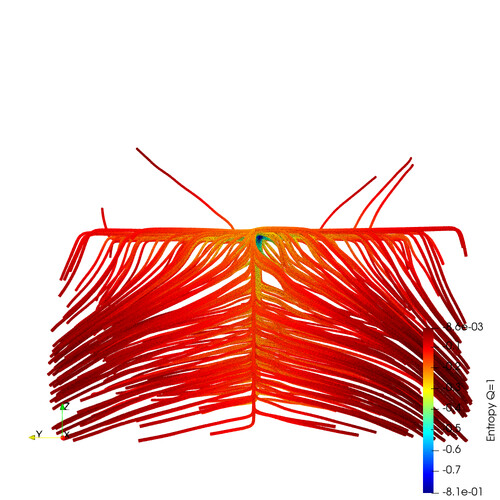}}%
\hfill
\subcaptionbox{}{\includegraphics[scale=\myscale]{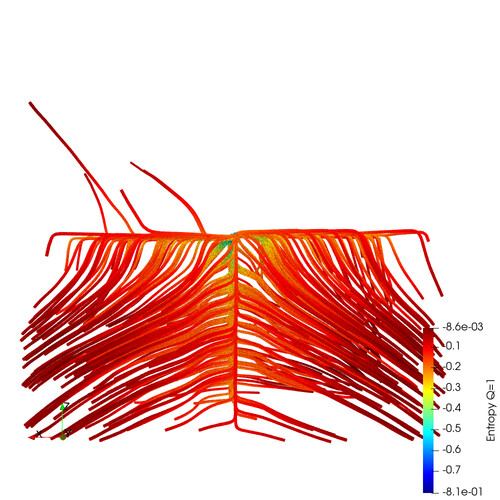}}%
\hfill
\caption{\label{path_scale_mattes}Paths of the gradient divergent algorithm along the MI of the same image with only scale transform and the ITK Mattes algorithm: (a) the paths emerging from the $z<0$ hemisphere on the back viewing from the $z$ axis (b) same as before from different angle (c) same as before from the $x$ axis (d) same as before from the $y$ axis}
\end{figure*}

The scale transform performs very well under ITK Mattes (Figure \ref{path_scale_mattes}) and we have all paths converging to the central point. One can only argue that the paths are not very direct but the true is that we can register the image from any scale transform and that is what matters.

\begin{figure*}
\centering
\subcaptionbox{}{\includegraphics[scale=\myscale]{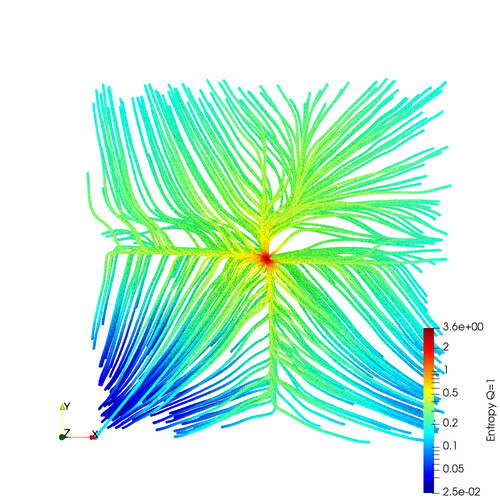}}%
\hfill
\subcaptionbox{}{\includegraphics[scale=\myscale]{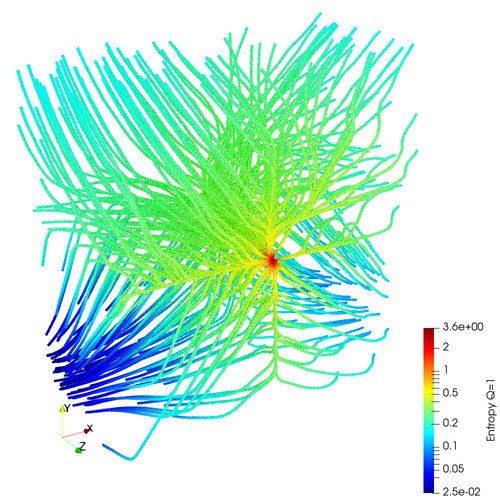}}%
\hfill
\subcaptionbox{}{\includegraphics[scale=\myscale]{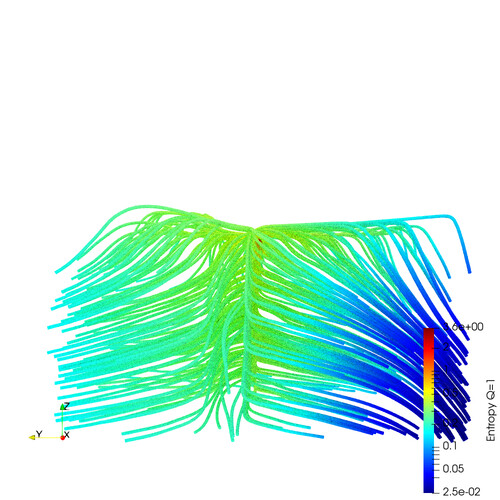}}%
\hfill
\subcaptionbox{}{\includegraphics[scale=\myscale]{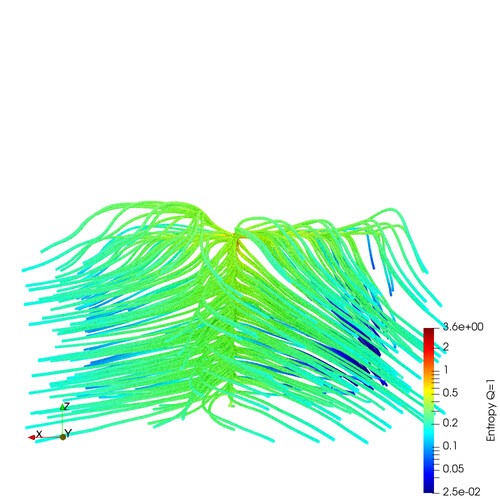}}%
\hfill
\caption{\label{path_scale_shannon}Paths of the gradient divergent algorithm along the MI of the same image with only scale transform and the MICUDA Shannon algorithm: (a) the paths emerging from the $z<0$ hemisphere on the back viewing from the $z$ axis (b) same as before from different angle (c) same as before from the $x$ axis (d) same as before from the $y$ axis}
\end{figure*}

The analysis of scale transform using the MICUDA Shannon function (Figure \ref{path_scale_shannon}) is very similar to the ITK Mattes one. The interesting aspect we can see is the upper right quadrant in the first image (a) have a different field than other quadrants. To understand this we need to understand the axis, this upper right quadrant is where $x>0$ and $y>0$, so we have a moving image that is scaled bigger than the fixed image in both $x$ and $y$ axis. One can reason that the algorithm scale down the bigger parameter first and then when we have some equal aspect ration on the scale ratios the algorithm will move to the central point. Such behavior appears in the lines flowing to a diagonal convergence and then flowing to the central point in the upper right quadrant.

On the other quadrants that we have any scale to bigger images (upper left and bottom right quadrant) we see a similar behavior were the paths flow first to eliminate this bigger scale parameter and then flow to the central point. This show some tendency of the MI function to first squeeze the moving image to a size similar to the fixed image, moving the parameters greater than zero to zero, and then growing the moving image in the axis it was smaller than the fixed or, in other words, moving the parameters lesser than zero to zero. In short MI with MICUDA Shannon seems to prioritize squeezing a bigger image than growing a small image.

\begin{figure*}
\centering
\subcaptionbox{}{\includegraphics[scale=\myhalfscale]{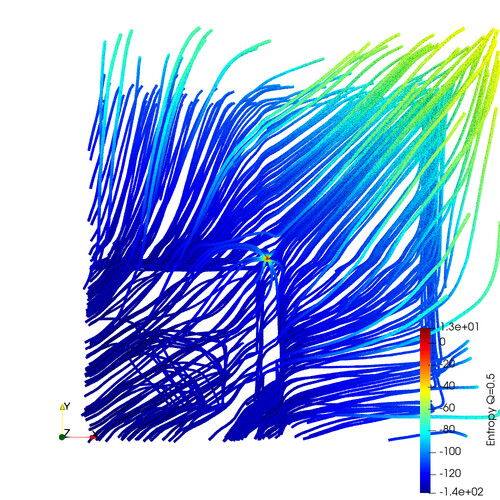}}%
\hfill
\subcaptionbox{}{\includegraphics[scale=\myhalfscale]{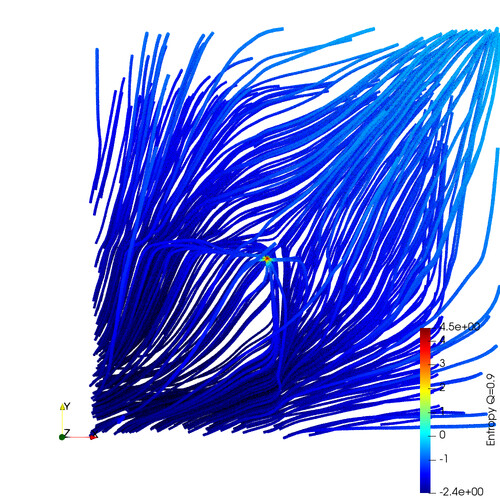}}%
\hfill
\subcaptionbox{}{\includegraphics[scale=\myhalfscale]{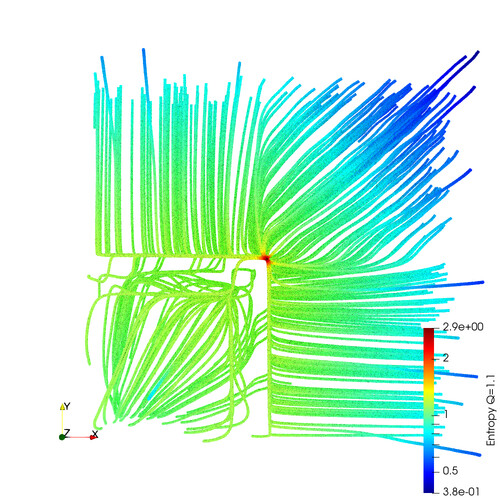}}%
\hfill
\subcaptionbox{}{\includegraphics[scale=\myhalfscale]{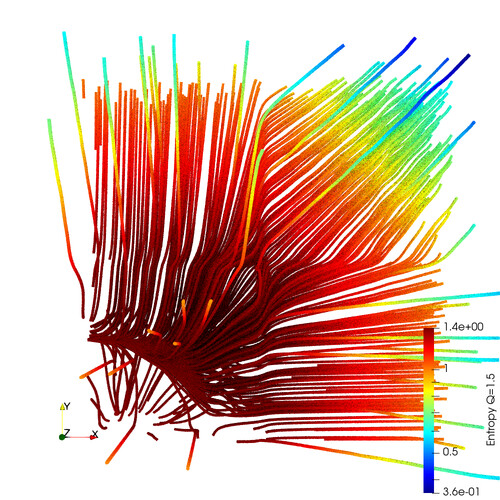}}%
\hfill\\
\subcaptionbox{}{\includegraphics[scale=\myhalfscale]{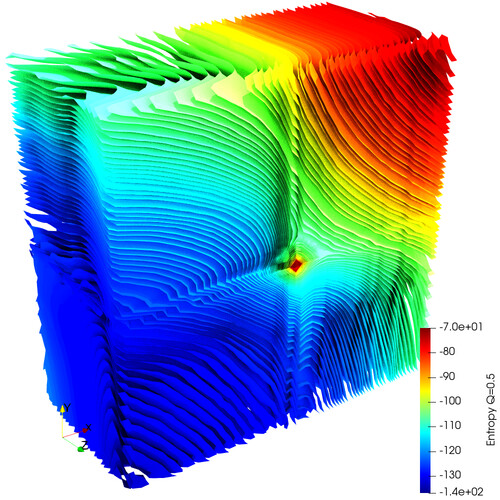}}%
\hfill
\subcaptionbox{}{\includegraphics[scale=\myhalfscale]{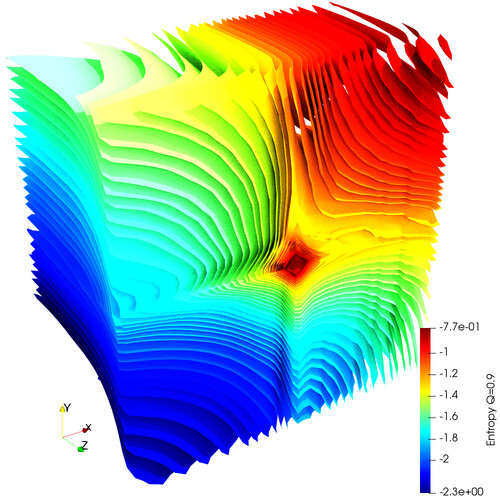}}%
\hfill
\subcaptionbox{}{\includegraphics[scale=\myhalfscale]{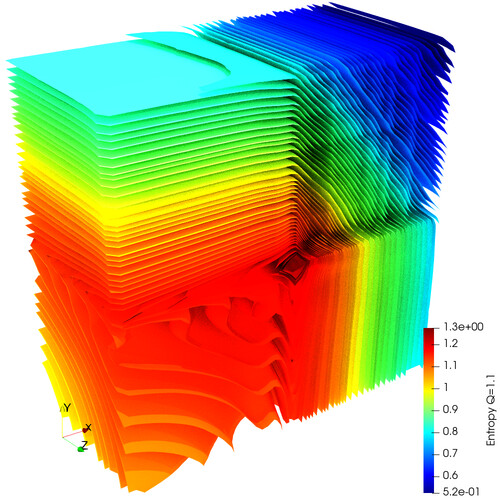}}%
\hfill
\subcaptionbox{}{\includegraphics[scale=\myhalfscale]{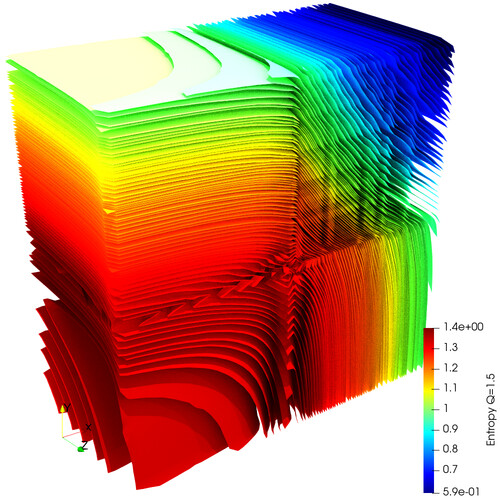}}%
\hfill
\caption{\label{path_scale_tsallis}Paths of the gradient divergent algorithm along the MI of the same image with only scale transform and the MICUDA Tsallis algorithm with different $q$ values: (a) $q=0.5$ (b) $q=0.9$ (c) $q=1.1$ (d) $q=1.5$ (e) Isosurfaces for $q=0.5$ (f) $q=0.9$ (g) $q=1.1$ (h) $q=1.5$}
\end{figure*}

Using the MICUDA Tsallis function we have very interesting but unfortunately not useful to registration. We see in Figure \ref{path_scale_tsallis} multiple graphs of different $q$ values. With $q<1.0$ (images a,b,e,f) we have the central point as a local maximum but we also have another local maximum and a very strong gradient field to the upper right quadrant. In those cases we can have a correct registration in only few cases that will direct to the central point. In any other case the algorithm will diverge in the sense to create a bigger moving image than the correct one. With $q>1.0$ (images c,d,g,h) the tendency changes and the algorithm will create a smaller image than the correct one. 

Only if we start from a point very near the central point (i.e. our moving image is pretty close to the fixed image in terms of size and won't need to be scaled or only need a very small scale) we will have a correct registration. In the cases were we really need to scale the image to register it we probably will have a wrong image in the end of the algorithm using the Tsallis entropy.

\subsection{Performance}
First of all it's not fair to compare some mathematical computation of GPU and CPU since they are very different in concept, with that in mind we note that our situation of many matrix and vectors computation is very beneficial to GPU usage.

We compared our new algorithm called MICUDA with the ITK Mattes algorithm. MICUDA runs mostly on GPU while ITK Mattes runs on CPU only. The hardware available for our tests was an Intel\textsuperscript{\textregistered} i7-2600K with 8 cores, were we also run our GPU, and a dual Intel\textsuperscript{\textregistered} Xeon 6130 Gold totaling 64 cores. The GPU used was a NVIDIA GTX 1060.

In Table \ref{performance} we have the numbers of our comparison, it's pretty clear that MICUDA using GPU is the winner with a speed up of about $140\times$ a high end 8 core computer (simulated with the Xeon). In the scenario of reviving an old computer the algorithm plays a major role, since with a simple GPU card upgrade we have speed ups of roughly $250-300\times$.

It also must be noted that the Xeon used is a very expensive hardware, specially if we compare to the price of the GPU card used. So in most realistic situations it may not be available to end users of a medical registration software.

\begin{table*}
\begin{tabular}{llllllll}
%\toprule
 & \multicolumn{3}{c}{ITK Mattes} & MICUDA & \multicolumn{3}{c}{Speed up (MICUDA vs)} \\ 
\cmidrule(lr){2-4} \cmidrule(lr){5-5} \cmidrule(lr){6-8}
Transform & Xeon $\times 8$ & Xeon $\times 64$ & i7-2600K & GTX 1060 & Xeon $\times 8$ & Xeon $\times 64$ & i7-2600K\\
\cmidrule(lr){1-5} \cmidrule(lr){6-8}
%\midrule
Translation & 0.8388 & 2.3060 & 0.4515 & 116.21 & 138.54 & 50.39 & 257.39\\ 
Rotation & 0.5862 & 1.6471 & 0.3097 & 86.30 & 147.22 & 52.40 & 278.66\\ 
Scale & 0.6665 & 1.8356 & 0.3076 & 99.46 & 149.23 & 54.18 & 323.34\\ 
\end{tabular}
\caption{Performance comparison between ITK Mattes and MICUDA algorithms in values processed per second.}
\label{performance}
\end{table*}

\section{Discussion}
Following the results on the former section we can see that the main problem in the ITK Mattes function is the translation transform. This can be solved very well using the MICUDA Tsallis algorithm allowing us to register images from all points in the space tested.

The rotation and scale transforms have a nice performance under the ITK Mattes algorithm and a similar performance can be reached using the MICUDA Shannon algorithm. Tsallis entropy have a very bad performance on those transforms and should not be used under the risk of producing very wrong registrations.

Since the computational cost of the MI function is mostly in the transform and histogram build and we can use the same histogram to produce the Shannon and Tsallis results the reasonable strategy is to use multiple algorithms in the gradient descent. In this way we can use the Tsallis in the translation gradients and Shannon in the rotation and scale gradients, having the best of both algorithms without much extra computational costs.

The ITK Mattes uses a random sampling of the image in the calculation and also uses a histogram with bins while MICUDA uses all the voxels and produces a full histogram. Further study is needed to understand if the bins and random sampling benefits the rotation and scale transforms and can improve Tsallis values. 

This study focus mostly on the capture range of the registration trying to fix the problem of registration not converging at all to the fixed image. More study is needed on the final accuracy of the registration in the very fine end or how close to the fixed image each algorithm can get. In some cases these final small tuning can be very difficult to reach since we start to see the effects of interpolation and our registered image may become offset by a few millimeters.

In the performance there's not much to say, the speed up of using a GPU card in this situation is huge. With a simple upgrade on actual or even older hardware we can have a better result with much less time, so it's a win in all aspects of the comparison.

\section{Acknowledgments}
Data were provided by the Human Connectome Project, WU-Minn Consortium (Principal Investigators: David Van Essen and Kamil Ugurbil; 1U54MH091657) funded by the 16 NIH Institutes and Centers that support the NIH Blueprint for Neuroscience Research; and by the McDonnell Center for Systems Neuroscience at Washington University.

\bibliography{bibliography}
\bibliographystyle{ieeetr}

\end{document}